%% file: universe-406834-eng-done.tex
\documentclass[universe,article,accept,moreauthors,pdftex,10pt,a4paper]{Definitions/mdpi} 

\firstpage{1} 
\pubvolume{4}
\issuenum{1}
\articlenumber{148}
\pubyear{2018}
\copyrightyear{2018}
\history{Received: 28 November 2018; Accepted: 13 December 2018; Published: 15 December 2018}
\updates{yes} 

\pdfoutput=1

\input macros

\usepackage{wrapfig}
\usepackage{amssymb}

\Title{Precision Determination of $\alpha_s$ from Lattice QCD}

\Author{Mattia Dalla Brida $^{1,2,}$*\orcidA{} on behalf of the ALPHA Collaboration}

\AuthorNames{Mattia Dalla Brida}

\address{%
$^{1}$ \quad Dipartimento di Fisica, Universit\`a di Milano-Bicocca, Piazza della Scienza 3, I-20126 Milano, Italy \\
$^{2}$ \quad INFN, Sezione di Milano-Bicocca, Piazza della Scienza 3, I-20126 Milano, Italy}

\corres{Correspondence: mattia.dallabrida@unimib.it}

\abstract{We present an overview of the recent lattice determination of the QCD coupling 
		  $\alpha_s$ by the ALPHA Collaboration. The computation is based on the non-perturbative
		  determination of the $\Lambda$-parameter of $\Nf=3$ QCD, and the perturbative matching 
		  of the $\Nf=3$ and $\Nf=5$ theories. The~final result: $\alpha_s(m_Z)=0.11852(84)$, 
		  reaches sub-percent accuracy.}


\begin{document}
\section{Introduction}

The strong coupling $\alpha_s$ is a fundamental parameter of QCD and
therefore of the Standard Model (SM) of Particle Physics.%
\footnote{As is custom in phenomenology, when we generically refer to 
		  the strong coupling $\alpha_s$ we actually refer to its value 
		  in the $\overline{\rm MS}$-scheme of dimensional regularization, 
		  evaluated at the $Z$-boson mass $m_Z$ i.e., $\alpha_s\equiv
		  \alpha_{\overline{\rm MS}}(m_Z)$.}
Its value affects the result of any perturbative calculation involving 
the strong interactions, and hence virtually all cross section calculations
for processes at the Large Hadron Collider. It also impacts considerations
on the stability of the electroweak vacuum, grand unification arguments, 
and searches of new coloured sectors.%
\footnote{For some recent reviews on the relevance of $\alpha_s$ to
		  phenomenology, we recommend the reader Refs.~\cite{Salam:2017qdl,
		  dEnterria:2018cye}.} %
At present, the PDG world average for the strong coupling has an uncertainly
$\delta\alpha_s/\alpha_s\approx0.9 \%$~\cite{Tanabashi:2018oca}. The coupling of 
QCD is thus by orders of magnitude, the least-precisely determined coupling among
those characterizing the (known) fundamental forces of nature. From the point of
view of phenomenology, the current uncertainty on $\alpha_s$ leads to relevant, 
i.e., 3--7\%, uncertainties in key Higgs processes such as $gg\to H$ and 
associated $H-t\bar{t}$ cross sections, and $H\to b\bar{b},c\bar{c},gg$ branching 
fractions. This uncertainty on $\alpha_s$ is moreover expected to dominate the 
parametric uncertainties in determinations of the top-quark mass and electroweak 
precision observables at future colliders  (see e.g.,~\cite{dEnterria:2018cye}
and references therein). A determination of $\alpha_s$ comfortably below the 
\emph{per-cent} level is called for in precision tests of the SM.

\begin{figure}[H]
\centering
	\includegraphics[scale=0.35]{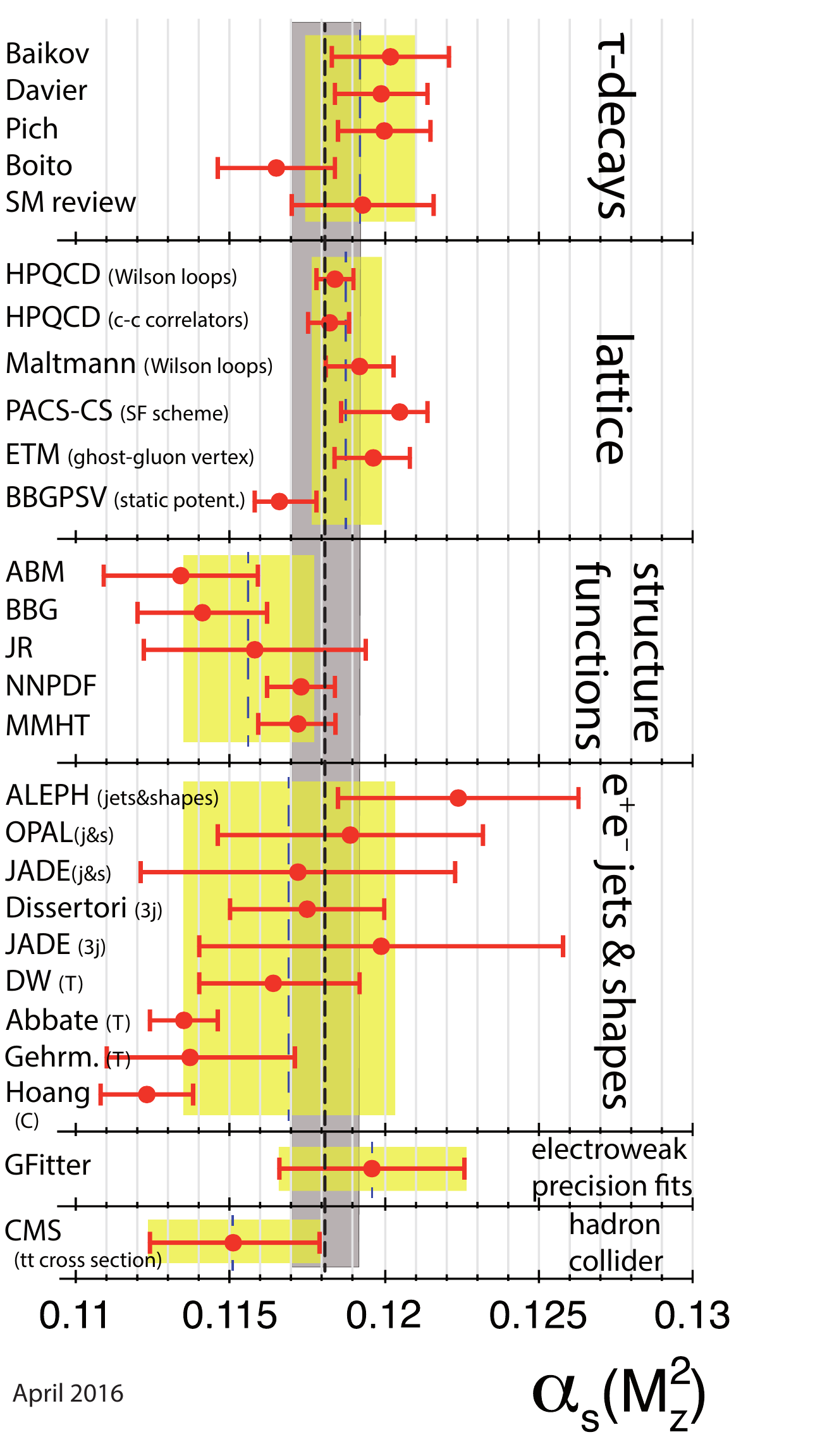}
	\caption{PDG summary of $\alpha_s$-determinations from six chosen 
			 sub-fields~\cite{Tanabashi:2018oca}. The yellow (light shaded) 
			 bands and dashed lines indicate the pre-average values of 
			 each sub-field. The dotted line and grey (dark shaded) band
			 represent the final world average vale of $\alpha_s$.}
	\label{fig:AlphaPDG}
\end{figure}

Currently, $\alpha_s$ is determined by the PDG by combining the results from
different sub-categories~\cite{Tanabashi:2018oca}. In short, the results 
from the different sub-fields are first pre-averaged, in such a way to take
into account the \emph{spread} among the different determinations (cf.~Figure~\ref{fig:AlphaPDG}). 
The sub-averages are then assumed to be independent and the final result for $\alpha_s$
is computed by \emph{$\chi^2$-averaging} these (cf.~Ref.~\cite{Tanabashi:2018oca}). 
The essential reason for the pre-averages is to obtain a more realistic estimate of 
$\alpha_s$ from the different sub-fields. Many determinations are in fact affected
by systematic uncertainties which are hard to quantify. It is thus likely that in 
some cases these might have been underestimated to some degree, causing the observed
spread. The situation is hence not quite satisfactory in general. For several determinations,
the attainable precision on $\alpha_s$ is likely limited by intrinsic sources of systematics, 
which in practice cannot be estimated to the desired level of precision. In order to achieve
a \emph{precise} and \emph{accurate} determination of the QCD coupling, it is thus mandatory
to develop methods where all sources of uncertainty can \emph{reliably} be kept under control. 
In this light, one might even object to the way $\alpha_s$ is currently determined. In the case 
of other fundamental parameters, only a small set of accurate procedures, if not a single one, 
is actually considered for their determination, while the others are intended as tests 
of the theory. As we shall discuss below, however, in the case of $\alpha_s$, finding such an 
accurate procedure is no simple task.

The outline of this contribution is the following. In the next section, we overview how
the strong coupling is generally determined in phenomenology, emphasizing in particular 
the main difficulties one encounters. We then move on to discuss determinations based on
lattice QCD methods, which are the focus of this presentation. We first discuss how a naive
approach has to face similar shortcomings as most phenomenological determinations, 
and see how these can be overcome by introducing finite-volume renormalization 
schemes and step-scaling techniques. The rest of the discussion is dedicated to presenting 
a specific lattice determination using these methods, recently completed by the ALPHA 
Collaboration~\cite{Brida:2016flw,DallaBrida:2016kgh,Bruno:2017gxd,DallaBrida:2018rfy}. 
The remarkable aspect of this result is that it achieves a sub-percent precision 
determination of $\alpha_s$ where the final error is dominated by \emph{statistical} 
rather than systematic uncertainties. We finally conclude discussing some future directions 
on how this result could be further~improved.

\section{Determinations of $\alpha_s$: General Considerations}
\unskip
\subsection{Phenomenological Determinations}
\label{subsec:AlphaFromPheno}

Any phenomenological determination of $\alpha_s$ is essentially based on the following 
idea (see e.g.,~Ref.~\cite{Salam:2017qdl}). One considers an experimental observable 
$\mathcal{O}(q)$ depending on an overall energy scale $q$; for simplicity, we ignore any 
dependence on additional kinematic variables. This observable is then compared with its 
theoretical prediction in terms of a perturbative expansion in 
$\alpha_s(\mu)\equiv\alpha_{\overline{\rm MS}}(\mu)$:
\begin{equation}
	\label{eq:FromOtoAlpha}
	\mathcal{O}_{\rm th}(q) = 
	\sum_{n=0}^N c_{n}(s) \alpha_s^n(\mu)
	+ {\rm O}(\alpha_s^{N+1}) 
	+{\rm O}\Big({\Lambda^p\over q^p}\Big),
	\quad
	\mu={q\over s},
\end{equation}
where the renormalization scale $\mu$ is set proportional to $q$ through the parameter 
$s$. The factors $c_n$ appearing in this equation are the coefficients of the perturbative 
series, which depend on the choice of renormalization scheme made for the coupling and on
the scale factor $s$. In practice, these are known up to some finite order $N$. Requiring: 
$\mathcal{O}_{\rm th}(q)=\mathcal{O}(q)$ fixes the value of the coupling $\alpha_s(\mu)$ 
up to some error, which comes from several different sources. First of all, the precision 
$\delta\mathcal{O}(q)$ to which the observable $\mathcal{O}(q)$ is known experimentally. 
It is the general situation that, when $q$ becomes large, and hence most sources of 
uncertainty become small (s.~below), the experimental errors in $\mathcal{O}(q)$ also 
get large. Second, is the effect of missing perturbative orders, i.e., the size of the 
${\rm O}(\alpha_s^{N+1})$ terms in Equation~(\ref{eq:FromOtoAlpha}). When the available 
measurements of $\mathcal{O}(q)$ allow it, this is normally accessed by studying the 
agreement between extractions of $\alpha_s(\mu)$ at different values of $q$, which are 
compared at some common reference scale through perturbative running. As we shall see in
a realistic case (cf.~Section~\ref{subsec:HighEnergy}), in order to reliably access this 
error, it is necessary to vary $q$ over a wide range, reaching up to large values (say 
O($100\,\GeV$)). In the situation where $q$ cannot really be varied, the 
${\rm O}(\alpha_s^{N+1})$ terms are usually estimated by studying the size of the known 
perturbative terms and their ``convergence''. Another strategy is to study the effect of 
varying $\mu$ rather than $q$, or equivalently the parameter $s$. Due~to the asymptotic 
character of the perturbative expansion, however, estimating the contribution of missing 
perturbative orders within perturbation theory itself is in principle an impossible task. 
In practice, this procedure is particularly questionable if $q$ is not really large and 
hence far from the region of asymptoticity. Another source of uncertainty in the extraction 
of $\alpha_s$ is the size of ``non-perturbative contributions''. These are formally represented 
in Equation~(\ref{eq:FromOtoAlpha}) by some power corrections to the perturbative expansion of 
${\rm O}({\Lambda^p/q^p})$, where $p>0$ and $\Lambda$ is some characteristic non-perturbative 
scale of QCD. In fact, our knowledge of the form of non-perturbative effects is rather limited. 
It is hence always debatable whether any model that tries to capture them is really adequate to
the describe the data within a given accuracy. A more systematic way to deal with the problem 
is again to study the extraction of $\alpha_s$ for a wide range of (large) $q$-values, and access
whether the use of perturbation theory alone gives consistent results. This in general means 
accessing whether any observed discrepancy is consistent with the effect of missing higher-order 
terms in the perturbative expansion. Finally, one should always keep in mind that any experimental 
observable $\mathcal{O}$ is in principle sensitive to all kinds of physics. The theoretical description
of $\mathcal{O}$ given by Equation~(\ref{eq:FromOtoAlpha}) might thus be missing, e.g., higher-order 
electroweak corrections, uncertainties associated with other SM parameters, or even effects
from yet unknown physics. 

In conclusion, a precise phenomenological determination of $\alpha_s$ has 
to face many challenges, and many determinations are limited in their precision by one or more of the
above issues. In the following sections, we would like to argue that extracting $\alpha_s$ using 
lattice QCD methods, on the other hand, allows in principle to circumvent many of these limitations. 

\subsection{Phenomenological Couplings and $\Lambda$-Parameters}
\label{subsec:PhenoCouplings}

For the discussion that will follow, it is useful to introduce the concepts of \emph{phenomenological 
couplings} and their \emph{$\Lambda$-parameters} (see e.g.,~Ref.~\cite{Sommer:2015kza} for a more 
extensive presentation). We can indeed associate to any short-distance observable $\Obs(q)$ a 
renormalized coupling $\alpha_\Obs(\mu)$ as ($c_n\equiv c_n(1)$):
\begin{equation}
	\alpha_\Obs(\mu)\equiv{\bar{g}^2_\Obs(\mu)\over4\pi} \equiv 
	{\Obs_{\rm th}(\mu) -c_0\over c_1} \overset{\mu\to\infty}{=} 
	\alpha_s(\mu) + {c_2\over c_1}\alpha_s^{2}(\mu)+\ldots\,.
\end{equation}

In this language, the extraction of $\alpha_s$ through $\Obs_{\rm th}$ is interpreted as the 
perturbative matching between the couplings $\alpha_\Obs$ and $\alpha_s$, where different 
observables $\Obs$ correspond to different \emph{renormalization schemes}.%
\footnote{We restrict our attention to mass-independent renormalization schemes which have 
		  simple renormalization group equations~\cite{Weinberg:1951ss}. These are obtained 
		  by considering the observable $\Obs$ for zero (renormalized) quark-masses.}
		 	
The value of the coupling $\bar{g}_\Obs(\mu)$ at any renormalization scale is in 
one-to-one correspondence with the $\Lambda$-parameter:
\begin{equation}
	\label{eq:LambdaParam}
	\Lambda_\Obs=
	\mu\times(b_0\bar{g}^2_\Obs(\mu))^{-b_1/(2b_0^2)} 
	e^{-1/(2b_0\bar{g}^2_\Obs(\mu))}
	\times
	\exp\bigg\{-\int_0^{\bar{g}_\Obs(\mu)} \rmd g \bigg[{1\over\beta_\Obs(g) } + 
	{1\over b_0 g^3} - {b_1\over b_0^2g}\bigg] \bigg\}.
\end{equation}

In this equation: $\beta_\Obs(g)\equiv[\mu\rmd\bar{g}_\Obs(\mu)/\rmd \mu]|_g$, is the 
well-known $\beta$-function, describing the dependence of the coupling on the renormalization
scale. In perturbation theory, we have:
\begin{equation}
	\label{eq:BetaFunctionPT}
	\beta_\Obs(g)\overset{g\to0}{\to}
	-b_0g^3-b_1g^5-b^\Obs_2g^7 +\ldots,
	\quad
	b_0=\big(11-\frac{2}{3}\Nf\big)(4\pi)^{-2},
	\quad
	b_1=\big(102-\frac{38}{3}\Nf\big)(4\pi)^{-4},
\end{equation}
where $\Nf$ is the number of (massless) quark-flavours considered. The coefficients $b_0,b_1$ are 
universal and shared by all mass-independent renormalization schemes; the scheme dependence only
enters through the higher-order coefficients $b_i^\Obs$, $i\geq 2$. It is easy to show that 
this implies a simple scheme dependence for the $\Lambda$-parameters. Specifically, we have that: 
${\Lambda_{\rm X}/\Lambda_{\rm Y}} = e^{c_{\rm XY}/(2b_0)}$, where $c_{\rm XY}$ is the one-loop 
matching coefficient between the schemes $X$ and $Y$, i.e., $\bar{g}_{\rm X}^2(\mu)=\bar{g}_{\rm Y}^2(\mu)
+ c_{\rm XY}\,\bar{g}_{\rm Y}^4(\mu) + {\rm }O(\bar{g}_{\rm Y}^6)$. 

\textls[-20]{We conclude with a couple of remarks. The $\Lambda$-parameters are \emph{exact} solutions of the 
Callan--Symanzik} equations~\cite{Callan:1970yg,Symanzik:1970rt,Symanzik:1971vw}. As such, they are 
renormalization group invariants ($\rmd\Lambda_\Obs/\rmd\mu=0$), and they are non-perturbatively 
defined if the coupling $\bar{g}_\Obs$ and its $\beta$-function are. In addition, as their
corresponding coupling, the $\Lambda$-parameters are defined for a \emph{fixed} number of 
quark-flavours $\Nf$. For the determination of $\alpha_s$, one needs $\Lambda^{\Nf=5}_{\overline{\rm MS}}$ 
(see e.g.,~Ref.~\cite{Tanabashi:2018oca}).

\subsection{Lattice QCD Determinations}

Lattice QCD provides a non-perturbative definition of QCD.%
\footnote{For a general introduction to lattice QCD, we suggest Refs.~\cite{Montvay:1994cy,Hernandez:2009zz}.
		  For a specific reference on determinations of $\alpha_s$ using lattice QCD, see 
		  instead \cite{Sommer:2015kza,Aoki:2016frl}.}	 
The theory is regularized by replacing the continuum Minkowskian space-time with a four-dimensional 
Euclidean space-time lattice, and by discretizing the QCD action, fields, and path integrals. 
The lattice spacing $a$, which separates two adjacent points of the lattice, provides an ultra-violet 
cutoff for the theory. Considering also a finite extent $L$ in all four space-time dimensions, the
number of degrees of freedom becomes finite and the theory is suitable for a numerical solution. 
In practice, the lattice path integrals representing field expectation values are evaluated stochastically
using Monte Carlo methods; any quantity measured in lattice QCD hence comes with some statistical 
uncertainty. Within the statistical precision, however, lattice QCD gives exact results for the given choice 
of lattice parameters. Physical results are then obtained by performing simulations at different values 
of the lattice spacing and lattice size, which allow the lattice results to be extrapolated to the continuum 
limit, $a\to0$, and infinite volume limit, $L\to\infty$. At present, lattice QCD actions normally contain
an unphysical number of quark-flavours $N_{\rm f}$, typically $N_{\rm f}=3,4$. In the following, we consider 
the situation where $N_{\rm f}=3$, which means that our action includes the up, down, and strange quarks. 
As custom in most lattice QCD set-ups, we moreover take the up and down quarks to be mass-degenerate.

In order to take the continuum limit, the theory must be properly renormalized. It is natural to renormalize 
lattice QCD in terms of hadronic inputs. More precisely, this means that the bare coupling $g_0$ and
the bare quark masses $m_{0,u}=m_{0,d}$, $m_{0,s}$, are fixed in such a way that lattice QCD reproduces
the experimental value of a number of hadronic quantities, as many as there are parameters. In our  
case, one could take, for instance: the proton mass $m_p$, and the dimensionless ratios $m_\pi/m_p$ and 
$m_K/m_p$, where $m_\pi$ and $m_K$ are the pion and kaon mass, respectively. In this example,
the proton mass is in fact used to set the value of the lattice spacing in physical units through: 
$a=(am_p)^{\rm lat}/m_p^{\rm exp}$, where $(am_p)^{\rm lat}$ is the proton mass in lattice units 
computed on the lattice, and $m_p^{\rm exp}$ is its experimental value. This given, we can express 
all other lattice quantities in physical units too.

Once lattice QCD has been renormalized in terms of a few low-energy inputs, any other quantity is 
in principle a \emph{prediction} of the theory. The value of the strong coupling $\alpha_s$, in particular, 
is now computable from first principles. A lattice QCD determination of $\alpha_s$ can be obtained
along the lines of what was discussed in Section \ref{subsec:AlphaFromPheno}, by simply taking for $\mathcal{O}(q)$
an observable "measured" in lattice QCD rather than experimentally. What are the advantages of this strategy? 
First of all, one has a lot of freedom in choosing $\mathcal{O}$. One can thus consider convenient observables
which have small uncertainties $\delta\mathcal{O}$. Being computed in lattice QCD, $\mathcal{O}$ is 
defined within QCD and QCD only. Hence, the corresponding theoretical description of Equation~(\ref{eq:FromOtoAlpha}) 
does not need to account for any contribution outside QCD. The~problem of disentangling the QCD contributions 
from experimentally measurable quantities is in fact confined to the hadronic quantities entering the 
renormalization of the theory. $\mathcal{O}(q)$ can in principle be computed fully non-perturbatively 
up to arbitrary large scales $q$. This allows one to have great control on both non-perturbative corrections
and missing perturbative orders in Equation~(\ref{eq:FromOtoAlpha}). Last but not least, lattice QCD 
is the \emph{only} known framework where no modelling of the hadronization of quarks and gluons into hadrons is
needed when comparing the measured observables with their theoretical (perturbative) predictions.

From this brief summary, it is clear that lattice QCD offers in principle many advantages for an accurate 
determination of the strong coupling. As we shall explain in the next subsection, this is indeed the case 
provided that some care is taken in choosing the observables $\mathcal{O}$.

\subsection{Finite-Volume Renormalization Schemes and Step-Scaling}
\label{subsec:StepScaling}

A naive application of lattice QCD methods to the determination of $\alpha_s$ has to face some technical 
limitations. Firstly, one must take into consideration that, in order to keep discretization effects under 
control in continuum extrapolations, the relevant energy scale of the observable $\mathcal{O}(q)$ must 
be well-below the ultra-violet cutoff set by the (inverse) lattice spacing. At the same time, the 
energy-scale $q$ must be well-above the typical non-perturbative scales of QCD, in order for the 
non-perturbative corrections and contributions from missing perturbative orders in Equation~(\ref{eq:FromOtoAlpha})
to be also under control when extracting $\alpha_s$. Furthermore, the space-time volume of the lattice 
simulations, controlled by $L$, must be large enough so that the observable $\mathcal{O}(q)$, as well as
the relevant hadronic observables needed to renormalize the lattice theory, are not affected by 
finite-volume effects. Putting all these constraints together, a safe extraction of $\alpha_s$ using lattice
QCD requires to satisfy e.g.,:
\begin{equation}
	\label{eq:WindowProblem}
	L^{-1}  \ll m_{\pi}  \ll m_{p} \ll q \ll a^{-1}.
\end{equation}

Considering that ideally one would want a factor 10 or so for each inequality appearing in Equation~(\ref{eq:WindowProblem}), 
we conclude that one should have at least $L/a={\rm O}(10^3)$. This however is already a factor 10 larger than what
is currently simulated in lattice QCD, where typically $L/a={\rm O}(10^2)$ with $a^{-1}\lesssim 5\,{\rm GeV}$. 
The only way one can proceed with this straightforward approach is thus to live with these lattices and make some 
severe compromises in Equation~(\ref{eq:WindowProblem}). In particular, the matching $\mathcal{O}_{\rm th}(q)=\mathcal{O}(q)$
may be performed at best at $q\approx a^{-1}={\rm O}(1\,\GeV)$. 

In fact, there is no real reason to compromise once a careful choice of observable $\mathcal{O}(q)$ is 
made. The basic problem of the naive strategy presented above is that it tries to accommodate on a single lattice, 
two widely separated energy scales: the non-perturbative scale of hadronic physics and the perturbative 
scale of weakly interacting quarks and gluons; this requires of course very fine lattices. 
However, it is not necessary to simulate all relevant energy scales within a single lattice QCD simulation: 
it is more natural to \emph{combine} different simulations, each one covering only a fraction of the energy
range of interest. The conditions for having systematic errors under control are then milder for each 
individual simulation. The way we can actually split the computation is to consider for $\mathcal{O}$ a 
\emph{finite-volume} observable $\mathcal{O}(q)\equiv\mathcal{O}(L^{-1})$, and introduce a corresponding 
\emph{finite-volume coupling} $\bar{g}^2_\Obs(\mu)$, $\mu=L^{-1}$, whose renormalization scale is given 
by the (inverse) size of the finite space-time volume of the system~\cite{Luscher:1991wu}. Said  
differently, we define a renormalized coupling through a finite-volume effect.  

The strategy to determine $\alpha_s$ using this class of observables is in short the following (see 
e.g.,~Ref.~\cite{Sommer:2015kza}). (1) The bare parameters of the theory are renormalized in terms of some 
hadronic inputs in a large physical volume with $m_\pi L\gg1$ and $am_p \ll1$. One then computes 
$\bar{g}^2_{\Obs}(\mu_{\rm had})$ at some low-energy scale $\mu_{\rm had}=L^{-1}_{\rm had}$, and 
establishes the exact relation between $\mu_{\rm had}$ and an hadronic scale, e.g., $\mu_{\rm had}/m_p=C$, 
where $C$ is a constant of O(1). No large scale separations are involved in this step, and one can satisfy 
\emph{simultaneously} the conditions: $a\mu_{\rm had}\ll1$, $am_p\ll 1$, and $m_\pi L\gg1$, which allow 
both discretization and finite-volume effects to be kept under control in this step; (2) One computes the
change in $\bar{g}^2_{\Obs}(\mu)$ as the scale $\mu$ is varied by a known factor. This is done through 
the step-scaling function (SSF) $\sigma_\Obs$ defined as:
\begin{equation}
	\sigma_\Obs(u)=\bar{g}_\Obs^2(\mu/2)|^{u=\bar{g}_\Obs^2(\mu)}_{\overline{m}(\mu)=0},
	\qquad
	\sigma_\Obs(u)=\lim_{a/L\to0} \Sigma_\Obs(a/L,u),
	\qquad
	\mu=L^{-1}.
\end{equation} 

\textls[-20]{The SSF is obtained by extrapolating to the continuum its lattice approximation $\Sigma_\Obs$. The~continuum} limit is performed at a fixed value of $\bar{g}_\Obs^2(\mu)=u$, setting the (renormalized)
quark-masses $\overline{m}(\mu)=0$. It~is important to note that the only condition that needs to be 
met for a safe continuum limit of the SSF is that $L/a\gg1$ i.e., the relevant energy scale $\mu\ll a^{-1}$; 
(3) By computing the SSF for a few steps, one is able to connect the low- and high-energy regimes of 
the theory. Starting from $\mu_{\rm had}=L^{-1}_{\rm had}$, in~particular, one can compute the value of
the coupling at $\mu_{\rm PT}=2^n\mu_{\rm had}\gg m_p$ by solving the recursion:
\begin{equation}
	\label{eq:StepScaling}
	\bar{g}_\Obs^2(\mu_{\rm had}) = u_{\rm had} = u_0 ,
	\qquad
	u_k = \sigma(u_{k+1}) = \bar{g}_\Obs^2(2^{k}\mu_{\rm had}),
	\qquad
	k = 0, 1, \ldots,n .
\end{equation}
(4) Given $\alpha_\Obs(\mu_{\rm PT})$ one option is to extract $\Lambda_\Obs/\mu_{\rm PT}$ using 
Equation~(\ref{eq:LambdaParam}) and the perturbative expression for $\beta_\Obs$ (cf.~Section~\ref{subsec:PhenoCouplings}). 
The perturbative relation: $\alpha_\Obs(\mu)=\alpha_{\overline{\rm MS}}(\mu)+ c_1\alpha^2_{\overline{\rm MS}}(\mu) +
c_2\alpha^3_{\overline{\rm MS}}(\mu)+{\rm O}(\alpha^4_{\overline{\rm MS}})$, then allows us to infer 
$\Lambda_{\overline{\rm MS}}/\mu_{\rm PT}$, from which $\alpha_{\overline{\rm MS}}(\mu_{\rm PT})$, with 
$\mu_{\rm PT}=2^n m_pC$, can be computed. Alternatively, one can directly match the couplings 
$\alpha_\Obs(\mu_{\rm PT})$ and $\alpha_{\overline{\rm MS}}(\mu_{\rm PT})$ using their perturbative~relation.

\section{The Schr\"odinger Functional and Finite-Volume Couplings}

There is in principle a lot of freedom in choosing the finite-volume coupling $\bar{g}_\Obs$.
From a practical point of view, however, there are a number of desirable properties that 
$\bar{g}_\Obs$ should have (see e.g.,~Ref.~\cite{Sommer:2015kza}). (1) The finite-volume 
coupling must be non-perturbatively defined and easily measurable within lattice QCD; (2) 
It should be computable in perturbation theory to at least NNLO with reasonable effort. 
Only in this case, we can expect to be able to extract $\alpha_s$ at high-energy with good
precision; (3) It should be gauge invariant, in order to avoid issues with Gribov copies once
studied non-perturbatively; (4) It should be quark mass-independent; (5) It must have small 
statistical errors when evaluated in lattice QCD and (6) have small lattice artefacts. Careful 
consideration about these points led to the definition of finite-volume couplings based on 
the Schr\"odinger functional (SF) of QCD~\cite{Luscher:1992an,Sint:1993un,Sint:1995rb}. 
Using a continuum language, the SF is formally defined by the partition function:
\begin{gather}
	\nonumber
	\mathcal{Z}_{\rm SF}[C,C']= 
	\int_{\rm SF\,bc}DA D\psi D\psibar\, e^{-S_{\rm QCD}},\\
	\nonumber
	A_k(x)|_{x_0=0}=C_k({\bf x}), 
	\quad
	A_k(x)|_{x_0=L}=C'_k({\bf x}),\\
	P_+\psi(x)|_{x_0=0}=0=P_-\psi|_{x_0=L},
	\quad
	\psibar(x)P_-|_{x_0=0}=0=\psibar P_+|_{x_0=L},
	\quad
	P_\pm=\frac{1}{2}(1\pm\dirac0).
\end{gather}

QCD is here considered in a finite Euclidean space-time with extent $L$ in all four dimensions. 
Periodic boundary conditions are assumed for the gauge field in the three spatial directions,
while, for the quark and anti-quark fields, they are periodic up to a phase~\cite{Sint:1995ch}.
In the temporal direction, the fields satisfy Dirichlet boundary conditions. The gauge 
field $A_\mu$ in particular is set equal to some ``classical'' field configurations $C_k,C_k'$ at  
Euclidean times $x_0=0,L$. Without entering into details, the SF has several nice properties. 
First of all, it is renormalized once the QCD parameters are renormalized. Secondly, perturbation 
theory with SF boundary conditions has been shown to be feasible up to at least NNLO~\cite{Narayanan:1995ex,
Bode:1998hd,Bode:1999sm,Brida:2013mva,DallaBrida:2016dai,DallaBrida:2017tru}. Furthermore, it allows 
lattice QCD simulations directly at zero quark-masses.

Even within the SF framework, however, it is not easy to find a single coupling definition that satisfies 
all the desired properties over the whole energy-range of interest. This, on the other hand, is not an
issue: one can consider different couplings with complementary properties in different parts of the 
energy-range. In the following, we introduce two such definitions, pointing out their pros and~cons.

\subsection{The SF Couplings}

A first family of SF based couplings can be defined in the following way; as is custom in the literature,
we shall refer to these simply as SF couplings. The boundary conditions for the gauge field are 
specified in terms of some spatially constant Abelian fields $C_k\equiv C_k(\eta,\nu)$ and 
$C'_k\equiv C'_k(\eta,\nu)$, which depend on two real parameters $\eta,\nu$; the exact
definition of these fields is not important here and can be found in e.g.,~Ref.~\cite{DallaBrida:2018rfy}. 
A family of renormalized couplings is then obtained as~\cite{Luscher:1992an,Luscher:1993gh,Sint:1995ch}:
\begin{equation}
	\bar{g}^2_{{\rm SF},\nu}(\mu) \equiv
	{k\over \partial_\eta \Gamma}\bigg|_{\eta=0},
	\qquad
	\Gamma\equiv-\ln\mathcal{Z}_{\rm SF}[C,C'],
	\qquad
	\mu=L^{-1},
\end{equation}
where $k$ is a constant that guarantees the normalization: $\bar{g}^2_{{\rm SF},\nu}=g_0^2+{\rm O}(g_0^4)$. 
Intuitively, these couplings are defined through the response of the system to an infinitesimal 
change of boundary conditions. Note that different $\nu$-values label different coupling definitions 
and thus schemes. A first result about the SF couplings is that their $\beta$-function is known to NNLO i.e.,
the coefficients $b_2^{{\rm SF},\nu}$ are known~\cite{Bode:1998hd,Bode:1999sm,DallaBrida:2018rfy}. It has 
been shown then that their statistical variance behaves like: ${\rm var}(\bar{g}^2_{{\rm SF},\nu})/
\bar{g}^4_{{\rm SF},\nu}\propto \bar{g}^4_{{\rm SF},\nu}$~\cite{deDivitiis:1994yz}. This means that, at 
high-energy, where $\bar{g}^2_{{\rm SF},\nu}$ is small, the statistical behaviour of these couplings is
improved. On the contrary, their statistical precision tends to deteriorate at low-energy. In addition, 
${\rm var}(\bar{g}^2_{{\rm SF},\nu})$ tends to be large in general, and increases $\propto L/a$ as the 
coupling is measured closer to the continuum limit, $a/L\to0$~\cite{deDivitiis:1994yz}. 
Fortunately, these couplings have small lattice artefacts, so that small values of $L/a$ are sufficient to 
perform reliable continuum extrapolations (cf.~Section~\ref{subsec:HighEnergy}). 

\subsection{The GF Coupling}

For the second coupling definition, we set $C_k=C'_k=0$, and introduce the Yang--Mills gradient flow 
(GF)~\cite{Narayanan:2006rf,Luscher:2010iy}:
\begin{gather}
	\nonumber
 	\partial_t B_\mu(t,x) = D_\nu G_{\nu\mu}(t,x),
 	\quad 
 	B_\mu(0,x)=A_\mu(x), \\
 	G_{\mu\nu} = \partial_\mu B_\nu - \partial_\nu B_\mu + [B_\mu,B_\nu],
 	\quad 
 	D_\mu = \partial_\mu + [B_\mu, \cdot\,].
 	\label{eq:GFequations}
\end{gather}

These equations define a flow gauge field $B_\mu(t,x)$ which depends on the flow time parameter 
$t\geq0$. At $t=0$, the flow field is given by the fundamental gauge field of the theory, but, as $t$ 
increases, it is driven towards some saddle point of the Yang--Mills action. The remarkable feature of 
the GF is that gauge-invariant fields made out of the flow field $B_\mu$ are \emph{renormalized}
quantities for $t>0$~\cite{Luscher:2011bx}. One can thus easily define a renormalized coupling by 
considering the simplest of these fields~\cite{Luscher:2010iy,Fodor:2012td,Fritzsch:2013je,DallaBrida:2016kgh}:
\begin{equation}
	\label{eq:GFcoupling}
	\bar{g}_{\rm GF}^2(\mu) = 
	\mathcal{N}^{-1}\,t^2\langle E_{\rm sp}(t,x)\rangle_{\rm SF}|_{x_0=L/2}^{\sqrt{8t} = 0.3 \times L},
	\quad
	E_{\rm sp}(t,x)=\frac{1}{4}G^a_{0k}(t,x)G^a_{0k}(t,x),
	\quad
	\mu=L^{-1}.
\end{equation}

The GF coupling, as we shall call it in the following, is here defined in terms of the spatial flow energy 
density $E_{\rm sp}(t,x)$ measured in the SF at $x_0=L/2$~\cite{Fritzsch:2013je}; the normalization constant 
$\mathcal{N}$ is chosen so that: $\bar{g}^2_{\rm GF}=g_0^2+{\rm O}(g_0^4)$. Note that, in order to have a 
proper finite volume coupling, this must depend only on a single scale given by $L$. For this reason, we expressed 
the flow time $t$ in terms of $L$ through the condition $\sqrt{8t}/L=0.3$; different values for this constant 
define \emph{different} renormalization schemes. The specific choice we made follows from a careful study of 
both the statistical properties and discretization effects of the different definitions~\cite{Fritzsch:2013je,
Ramos:2015baa,DallaBrida:2016kgh}. Similar investigations made us prefer the spatial part of the flow energy
density measured in the middle of the space-time volume over the full energy density.
 
A general nice property of GF based observables is their statistical precision. For the GF coupling, 
${\rm var}(\bar{g}^2_{\rm GF})$ is indeed typically small and essentially constant as $a/L\to0$. 
In addition, one can show that ${\rm var}(\overline{g}^2_{\rm GF})\,/\,\overline{g}_{\rm GF}^4\propto {\rm const.}$, 
which makes this coupling compelling for low-energy studies~\cite{Brida:2014joa}. A first drawback 
of $\bar{g}_{\rm GF}$ is that, at present, the available perturbative information is limited. The 
$\beta$-function at NNLO is only available for the pure SU(3) Yang--Mills theory~\cite{DallaBrida:2016dai,
DallaBrida:2017tru}. Moreover, current experience seems to indicate that, in general, the GF coupling 
tends to have largish lattice artefacts, which require largish lattice resolutions in order to have safe 
continuum extrapolations (cf.~Section~\ref{subsec:LowEnergy}).

\section{\boldmath{$\alpha_s$} from the Femto-Universe}

In this section, we present our computation of $\alpha_s$ based on finite-volume couplings. More details can 
be found in the original references~\cite{Brida:2016flw,DallaBrida:2016kgh,Bruno:2017gxd,DallaBrida:2018rfy}. 
The computation relies on the determination of $\Lambda^{\Nf=5}_{\overline{\rm MS}}$, which we divided 
into several factors, each one corresponding to a different step we describe below:
\begin{equation}
	\label{eq:MasterFormula}
	\Lambda^{\Nf=5}_{\overline{\rm MS}} =
	{\Lambda^{\Nf=5}_{\overline{\rm MS}} \over \Lambda^{\Nf=3}_{\overline{\rm MS}} }
	\times
	{\Lambda^{\Nf=3}_{\overline{\rm MS}} \over \mu_{0}}
	\times 
	{\mu_0 \over \mu_{\rm had}}
	\times 
	\Bigg[
	{\mu_{\rm had}\over\mu^*_{\rm ref}}
	\times {\mu^*_{\rm ref}\over f_{\pi K}}
	\times
	f^{\rm (PDG)}_{\pi K}
	\Bigg].
\end{equation}

\subsection{Determination of ${\Lambda^{\Nf=3}_{\overline{\rm MS}}/\mu_{0}}$ and 
			the Accuracy of Perturbation Theory at High-Energy}
\label{subsec:HighEnergy}

\begin{figure}[H]
	\centering
	\includegraphics[scale=0.5]{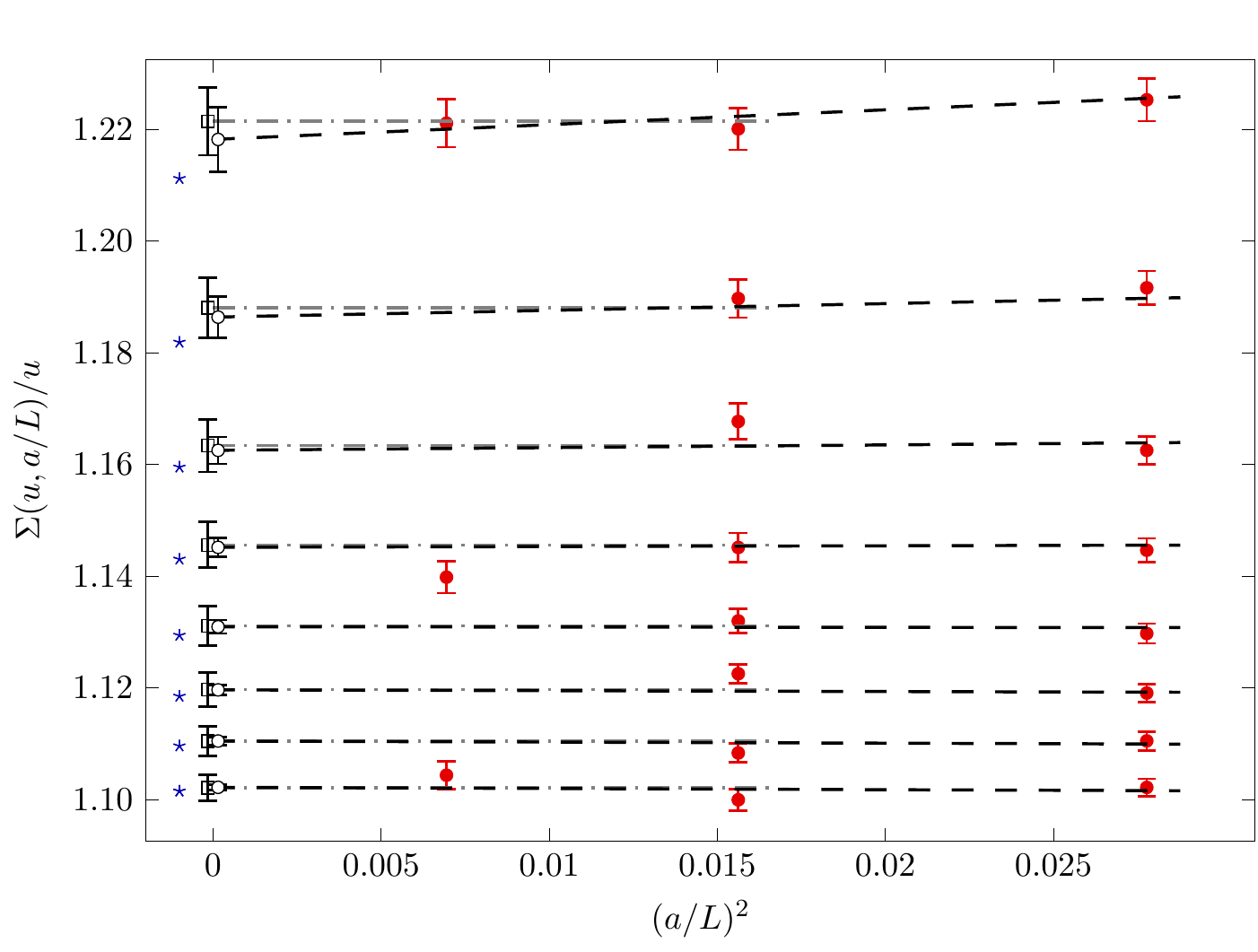}
	\quad
	\includegraphics[scale=0.59]{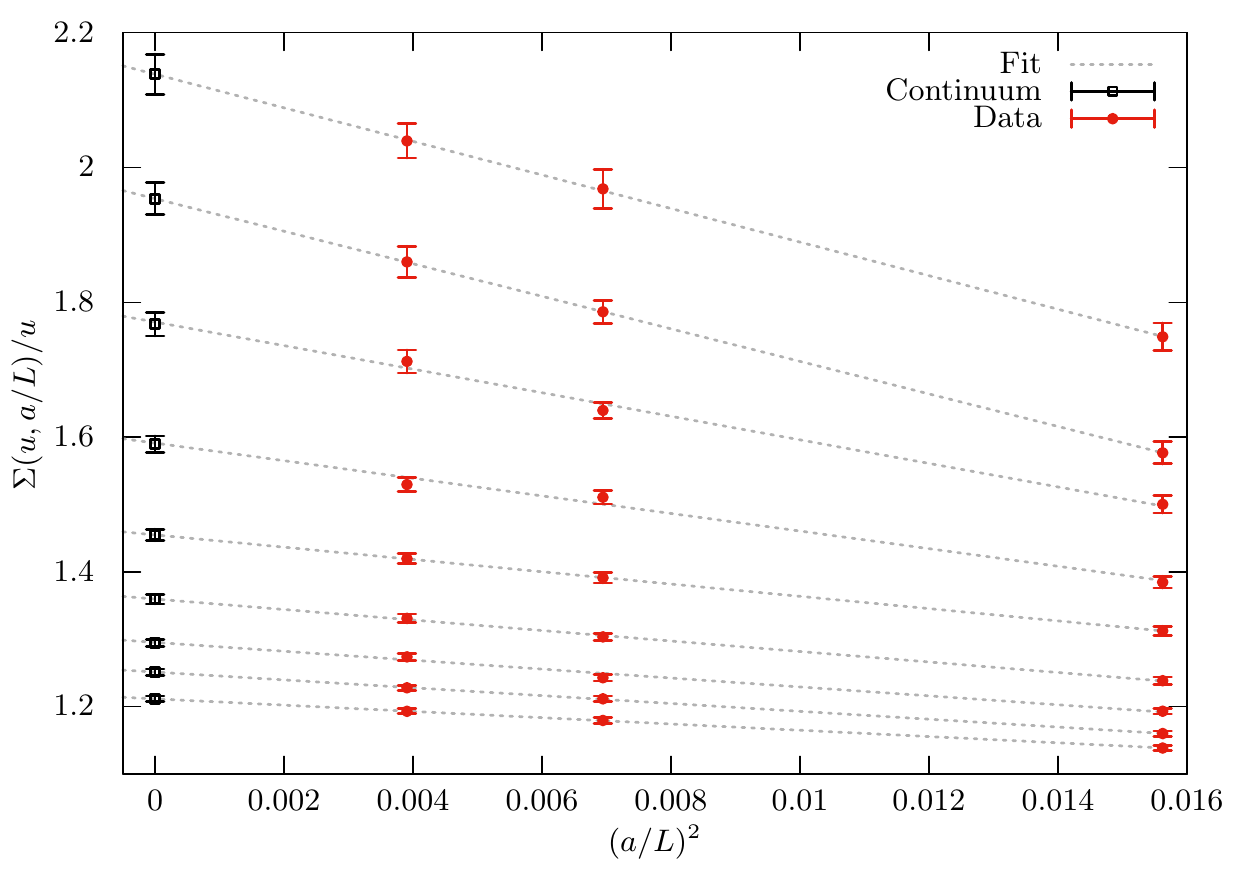}
	\caption{Continuum extrapolations of the lattice SSF of 
			 $\bar{g}^2_{{\rm SF},\nu=0}$ (left panel) and 
			 $\bar{g}^2_{\rm GF}$ (right panel), for different values
			 of the couplings. The lattice data is shown in red while
			 the black points are the continuum extrapolated results.
			 For the details on the extrapolations, we refer to
			 \cite{DallaBrida:2018rfy} and \cite{DallaBrida:2016kgh}, 
			 respectively.}
	\label{fig:LatticeSSFs}
\end{figure}

The determination of $\Lambda^{\Nf=5}_{\overline{\rm MS}}$ passes through the one of 
$\Lambda^{\Nf=3}_{\overline{\rm MS}}$, which we can compute non-perturbatively through
lattice QCD simulations in the $\Nf=3$ theory. For the computation of $\Lambda^{\Nf=3}_{\overline{\rm MS}}$, 
we begin by introducing a technical high-energy scale, $\mu_0$. This is 
done implicitly by choosing a relatively small value of the SF coupling at $\nu=0$: 
$\bar{g}^2_{{\rm SF},\nu=0}(\mu_0)=2.012$, with $\mu_0=L_0^{-1}$. The~lattice SSFs of 
the SF couplings are then determined non-perturbatively for several different values of 
the couplings, $\bar{g}^2_{{\rm SF},\nu}\approx 1-2$ ($|\nu|\approx0.1-1$), and for different 
lattice resolutions $L/a=4-12$~\cite{Brida:2016flw,DallaBrida:2018rfy}. The~chosen coupling 
range covers, in fact, five steps of step-scaling, i.e., a factor $32$ in energy. As expected, 
discretization errors in the lattice SSFs are very mild for the SF couplings, which allows us to 
obtain precise and robust continuum extrapolations. This can be appreciated in the left panel of 
Figure~\ref{fig:LatticeSSFs}, where the continuum  extrapolations of the lattice SSF for the 
$\nu=0$ case are shown. Employing the same set of simulations, we also determine the values of 
the couplings $\bar{g}^2_{{\rm SF},\nu}(\mu_0)$ with $\nu\neq0$. Given these and the results 
for the continuum SSFs, we can infer $\bar{g}^2_{{\rm SF},\nu}(\mu_n)$ at the scales $\mu_n=2^n\mu_0$, 
with $n=1,\ldots,5$ (cf.~Equation~(\ref{eq:StepScaling})). In conjunction with the NNLO $\beta$-function's, 
we can then use any of these coupling values in Equation~(\ref{eq:LambdaParam}) to extract 
$\Lambda^{\Nf=3}_{{\rm SF},\nu}/\mu_0$ (cf.~Section~\ref{subsec:StepScaling}). In fact, using 
the exact relation between the $\Lambda^{\Nf=3}_{{\rm SF},\nu}$'s for different $\nu$~\cite{DallaBrida:2018rfy}, 
it is convenient to convert all determinations to a common scheme: $\Lambda\equiv\Lambda^{\Nf=3}_{{\rm SF},\nu=0}$.

\begin{figure}[H]
	\centering
	\includegraphics[scale=0.45]{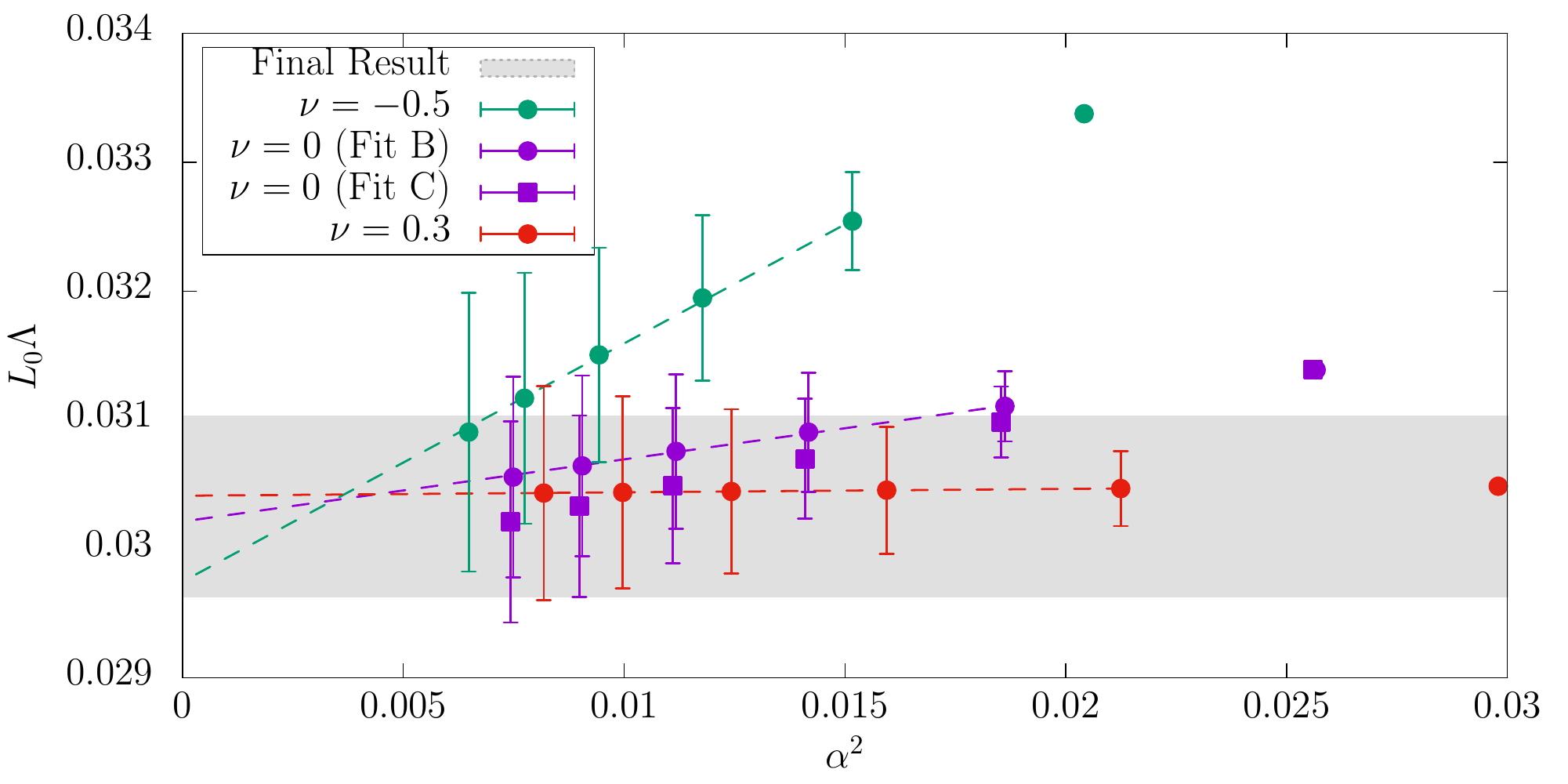}
	\caption{Determination of $L_0\Lambda\equiv\Lambda^{\Nf=3}_{{\rm SF},\nu=0}/\mu_0$ 
		at different values of $\alpha\equiv\alpha_{{\rm SF},\nu}(\mu_n)$. We compare
		the extraction in different schemes ($\nu=-0.5,0,0.3)$, and show a comparison
		with the final result Equation~(\ref{eq:LambdaL0}).  As one can see, when the
		extraction is performed at high enough energies $(\alpha\approx0.1)$, all~schemes
		agree.}
	\label{fig:Accuracy}
\end{figure}

The outcome of this procedure is shown in Figure~\ref{fig:Accuracy}, where the results for 
$\Lambda/\mu_0$ obtained from different $\alpha_{{\rm SF},\nu}(\mu_n)$ are displayed. 
Having extracted $\Lambda/\mu_0$ by truncating the $\beta$-function's in Equation~(\ref{eq:LambdaParam})
to NNLO, we expect the results obtained at different values of $\mu_n$ and for different $\nu$'s 
to differ by ${\rm O}(\alpha^2)$ corrections, as $\alpha\to0$~\cite{Brida:2016flw,DallaBrida:2018rfy}. 
This expectation is well-verified by the data. This may be taken as evidence that, in 
this range of couplings, non-perturbative corrections are negligible within our precision, and 
we are only sensitive to higher-order corrections in perturbation theory. It is moreover important
to note that, while there are schemes ($\nu=0.3$) where the corrections due to missing perturbative
orders are practically absent, there are other cases ($\nu=-0.5$) where these are quite significant. 
In particular, only when the couplings reach a value of $\alpha\approx0.1$, the differences between 
the results from different schemes become irrelevant within the statistical errors. There, we can safely
quote~\cite{DallaBrida:2018rfy}:
\begin{equation}
	\label{eq:LambdaL0}
	\Lambda/\mu_0=0.0303(7)
	\quad
	\Rightarrow
	\quad
	\Lambda_{\overline{\rm MS}}^{\Nf=3}/\mu_0=0.0791(19).
\end{equation}

This realistic example should be taken as a general warning for $\alpha_s$-determinations.
It is clear that unless one can test the accuracy of perturbation theory over a wide range of 
energies reaching up to high-energy (here we reach $\alpha\approx0.1$), it is not possible 
to estimate reliably the systematic errors coming from non-perturbative corrections and 
the truncation of the perturbative expansion. 

\vspace{-6pt}

\subsection{Low-Energy Running and Determination of $\mu_0/\mu_{\rm had}$}
\label{subsec:LowEnergy}

\textls[-25]{In order to express $\Lambda_{\overline{\rm MS}}^{\Nf=3}$ in physical units,
we must link the technical scale $\mu_0$ with some experimentally} accessible 
hadronic quantity. In order to keep systematic errors under control, we shall 
first relate $\mu_0$ with another finite-volume scale $\mu_{\rm had}={\rm O}(100\,\MeV)$, 
and then relate this with some hadronic quantity. 
We do this in the following way. Through some dedicated simulations, we first relate 
$\bar{g}^2_{{\rm SF},\nu=0}(\mu_0)$ with the GF coupling of Equation~(\ref{eq:GFcoupling}) at 
the scale $\mu_0/2$, obtaining: $\bar{g}^2_{\rm GF}(\mu_0/2)=2.6723(64)$~\cite{DallaBrida:2016kgh}. 
Then, similarly to what we did for the SF couplings, we compute the lattice SSF of the
GF coupling for several values of $\bar{g}^2_{\rm GF}\approx2-6.5$, and lattice sizes, 
$L/a=8-16$. The lattice SSFs are then extrapolated to the continuum, and a 
parametrization of the continuum SSF is obtained~\cite{DallaBrida:2016kgh}. 
In this case, the continuum extrapolations are more delicate than in the case of the SF couplings, 
due to significantly larger discretization errors (cf.~right panel of 
Figure~\ref{fig:LatticeSSFs}). Nonetheless, a~very good precision is attained for 
the final continuum results thanks to the high statistical accuracy of the GF coupling.

Having the (continuum) SSF, one can determine the \emph{non-perturbative} $\beta$-function of 
the coupling. The exact relation between the two is obtained by noticing that:
\begin{equation}
	\label{eq:ScaleRatio}
	\ln{\mu_2\over \mu_1} = 
	\int_{\bar{g}_\Obs(\mu_1)}^{\bar{g}_\Obs(\mu_2)} 
	{\rmd g \over \beta_\Obs(g)}
	\quad
	\Rightarrow
	\quad
	\log 2 = 
	-\int_{\sqrt{u}}^{\sqrt{\sigma_\Obs(u)}} {\rmd g \over \beta_\Obs(g)}
	\quad
	\text{where}
	\quad
	u=\bar{g}^2_\Obs(\mu).
\end{equation}

\begin{figure}[H]
	\centering
	\includegraphics[scale=0.58]{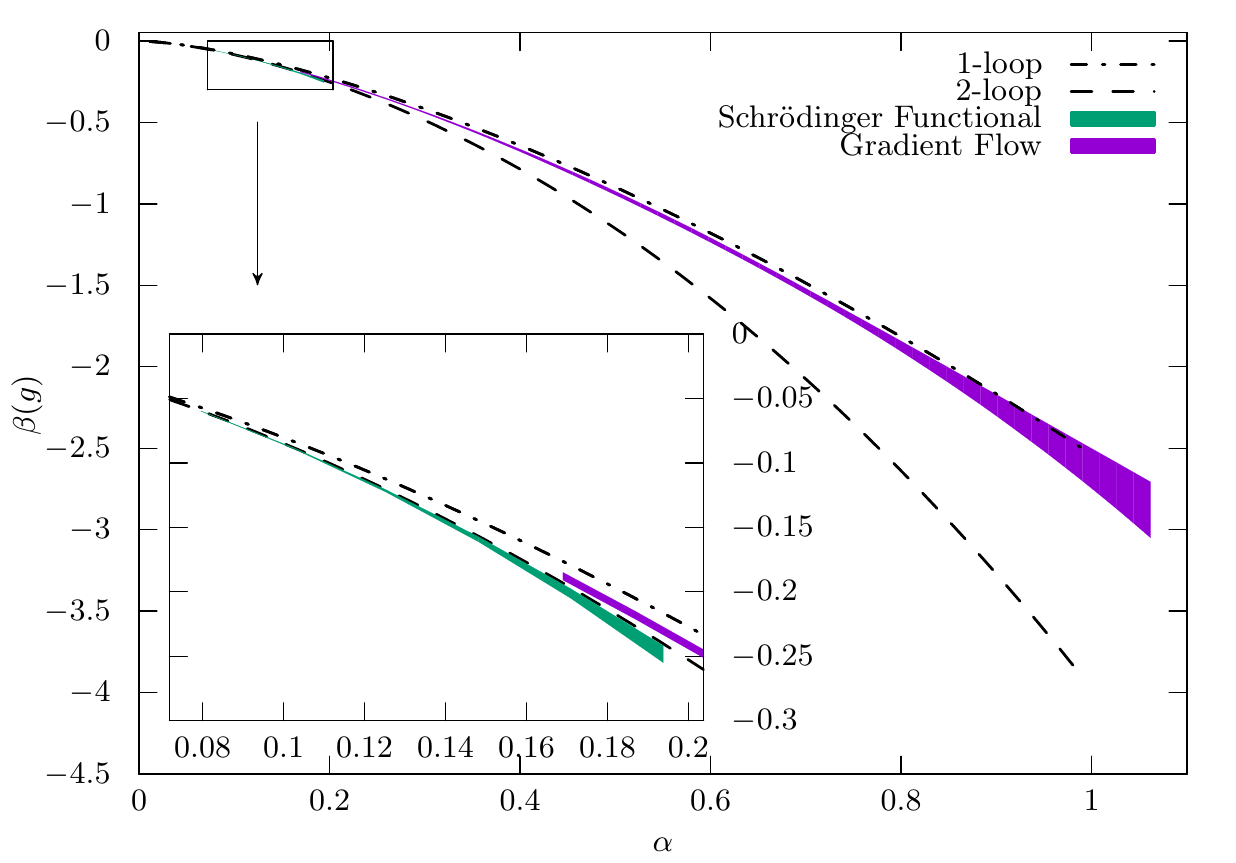}
	\quad
	\includegraphics[scale=0.6]{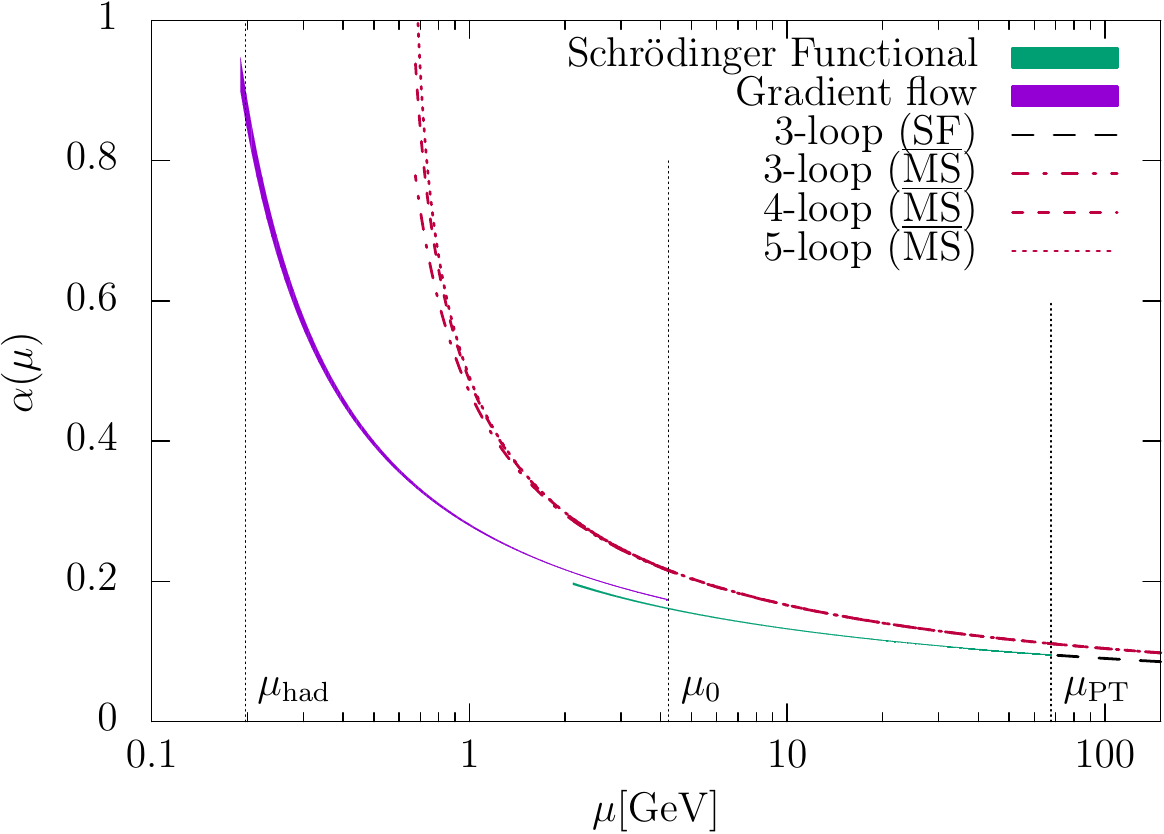}
	\caption{(\textbf{Left}): Non-perturbative $\beta$-functions in the GF and SF $(\nu=0)$ schemes
		determined in refs.~\cite{DallaBrida:2016kgh} and \cite{Brida:2016flw,DallaBrida:2018rfy},
		respectively. (\textbf{Right}): Running couplings of $\Nf=3$ QCD obtained from 
		$\Lambda^{\Nf=3}_{\overline{\rm MS}}$ by integrating the non-perturbative
		$\beta$-functions~\cite{Bruno:2017gxd}.}	
	\label{fig:BetaAndRunning}
\end{figure}

The results for the non-perturbative $\beta$-function of the GF coupling are 
shown in the left panel of Figure~\ref{fig:BetaAndRunning}, together with 
the results for the non-perturbative $\beta$-function of the SF coupling
analogously obtained; a comparison with their (universal) LO and NLO predictions
is also shown (cf.~Equation~(\ref{eq:BetaFunctionPT})). Observe the peculiar behaviour 
of the non-perturbative $\beta$-function of the GF coupling, which lies very close 
to its LO result even at large values of the coupling, where $\alpha\approx 1$. Note, 
however, that, for most of the coupling range, we studied, the deviation from LO 
perturbation theory is statistically significant~\cite{DallaBrida:2016kgh}. Only at 
values of $\alpha\approx0.2$ do the non-perturbative results then start to approach their 
NLO prediction.

The $\beta$-function allows us to compute the ratio of energy 
scales corresponding to any two values of the coupling (cf.~Equation~(\ref{eq:ScaleRatio})). 
Defining the technical scale $\mu_{\rm had}$ through a relatively large value of the GF 
coupling: $\bar{g}^2_{\rm GF}(\mu_{\rm had})=11.31$, Equation~(\ref{eq:ScaleRatio})
and the non-perturbative results for $\beta_{\rm GF}$ give us~\cite{DallaBrida:2016kgh,Bruno:2017gxd}:
\begin{equation}
{\mu_0/ \mu_{\rm had}}=21.86(42)
\quad
\Rightarrow
\quad
{\Lambda^{\Nf=3}_{\overline{\rm MS}}/\mu_{\rm had} } = 1.729(57).
\end{equation}
\vspace{-12pt}

\subsection{Hadronic Matching and $\Lambda_{\overline{\rm MS}}^{\Nf=3}$} 
\label{subsec:HadronicMatching}

\textls[-20]{All that is left to do to determine $\Lambda_{\overline{\rm MS}}^{\Nf=3}$ 
is to relate the finite-volume scale $\mu_{\rm had}$ to some experimentally }
accessible quantity. This is best done by introducing a convenient intermediate 
reference scale, $\mu_{\rm ref}=1/\sqrt{8t_0}$~\cite{Bruno:2017gxd}. The quantity 
$t_0$ corresponds to a specific flow time, which is implicitly defined by the flow 
energy density through the equation 
(cf.~Equations~(\ref{eq:GFequations})-(\ref{eq:GFcoupling}))~\cite{Luscher:2010iy}: 
\begin{equation}
	t^2_0\,\langle E(t_0,x) \rangle = 0.3,
	\qquad
	E(t,x)=\frac{1}{4}G^a_{\mu\nu}(t,x)G^a_{\mu\nu}(t,x).
\end{equation}

Note that differently from the case of the GF coupling, Equation~(\ref{eq:GFcoupling}), 
the expectation value appearing in this equation is the one of the theory in infinite 
space-time volume. Some of the nice features of the scale $\mu_{\rm ref}$ are that
it can be determined very accurately in lattice QCD simulations and at a very modest 
computational effort. Moreover, theoretical arguments and numerical evidence 
show that it depends very little on the value of the quark masses which 
are simulated. This is a great advantage in performing extrapolations to physical quark
masses when these are necessary. Of course, $\mu_{\rm ref}$ is not measurable in 
experiments, and its value in physical units must be determined by connecting it, through 
lattice QCD, to some experimentally accessible quantity. Once the value of $\mu_{\rm ref}$
in physical units is known, it is generally more convenient to express lattice quantities in terms
of $\mu_{\rm ref}$ rather than directly in terms of an hadronic quantity (see e.g.,~Ref.~\cite{Sommer:2014mea}). 
This is because usually the former can be more simply and accurately determined. 

The value of $\mu_{\rm ref}$ in physical units was obtained in Ref.~\cite{Bruno:2016plf}, to which we 
refer for any detail. In~short, this was computed employing an extensive set of large volume simulations 
of $\Nf=3$ QCD~\cite{Bruno:2014jqa}, and a precise renormalization programme~\cite{Brida:2016rmy,
DallaBrida:2018tpn}, which allowed to determine, in the continuum, infinite volume limit, 
and for physical up, down and strange quark masses, the ratio: $f_{\pi K}/\mu_{\rm ref}$, 
where $f_{\pi K}=\frac{2}{3}(f_K+\frac{1}{2}f_\pi)$ is a combination of pion and kaon 
decay constants. Using the PDG value for $f_{\pi K}$, one then arrives at: $\mu_{\rm ref}=478(6)\,\MeV$. 
In~fact, for technical reasons, rather than using the value of $\mu_{\rm ref}$ at physical quark 
masses, we shall use $\mu_{\rm ref}$ evaluated for \emph{equal} up, down, and strange quark 
masses, close to their average physical value~\cite{Bruno:2017gxd}; we indicate this "new" scale as 
$\mu^*_{\rm ref}=1/\sqrt{8t^*_0}$. Its value in physical units can be determined once again through 
lattice QCD simulations by relating $\mu^*_{\rm ref}$ to $f_{\pi K}$ or $\mu_{\rm ref}$. Doing this one 
obtains: $\mu_{\rm ref}^*=478(7)\,\MeV$~\cite{Bruno:2014jqa}, which actually shows how $\mu_{\rm ref}$ 
depends very little on the value of the quark masses. Having this result, we can obtain 
$\Lambda_{\overline{\rm MS}}^{\Nf=3}$ in physical units by computing the ratio $\mu^*_{\rm ref}/\mu_{\rm had}$ 
(cf.~Equation~(\ref{eq:MasterFormula})). Given the large volume results for $a\mu^*_{\rm ref}$ at several values 
of the lattice spacing $a$~\cite{Bruno:2016plf}, this can be determined with modest computational effort 
from a small set of lattice QCD simulations of the SF~\cite{Bruno:2017gxd}. The result we obtain is:
\begin{equation}
	\label{eq:LambdaNf3}
	\mu^*_{\rm ref}/\mu_{\rm had}= 2.428(18)
	\quad
	\Rightarrow
	\quad
	\Lambda^{\Nf=3}_{\overline{\rm MS}} = 341(12)\,\MeV.
\end{equation}

From $\Lambda^{\Nf=3}_{\overline{\rm MS}}$ and the non-perturbative $\beta$-functions of the SF
and GF couplings, one can reconstruct the non-perturbative running of the couplings
over the whole range of energy we covered, which goes from $\mu_{\rm had}\approx 200\,\MeV$
to $\mu_{\rm PT}=16\mu_0\approx 70\,\GeV$. The result is shown in the right panel of 
Figure~\ref{fig:BetaAndRunning}.

\subsection{Perturbative Decoupling and $\alpha_s$}

\textls[-25]{To arrive at $\alpha_s(m_Z)$, one needs $\Lambda_{\overline{\rm MS}}^{\Nf=5}$. The way we obtain 
this is to take as an input the non-perturbatively} determined $\Lambda_{\overline{\rm MS}}^{\Nf=3}$
of Equation~(\ref{eq:LambdaNf3}), and use perturbation theory to compute the ratio $\Lambda_{\overline{\rm MS}}^{\Nf=5}/
\Lambda_{\overline{\rm MS}}^{\Nf=3}$~\cite{Bruno:2017gxd}. More precisely, assuming a \emph{perturbative
decoupling} of the charm and bottom quarks, the couplings of QCD with $\Nf=5$ and $\Nf=3$ quark-flavours
are matched using perturbation theory at the charm and bottom quark-mass thresholds. This matching then
translates into a perturbative relation between the $\Lambda$-parameters of the two theories 
(see refs.~\cite{Athenodorou:2018wpk,Bruno:2014ufa} for the details). The reliability and accuracy of 
perturbation theory in this step has been recently addressed in a systematic way in Ref.~\cite{Athenodorou:2018wpk}. 
The conclusion of this study is that perturbation theory is expected to be reliable in predicting this
ratio of $\Lambda$-parameters to a \emph{sub-percent} accuracy. Since our determination of $\Lambda_{\overline{\rm MS}}^{\Nf=3}$
has a precision of about 3\%, we expect that the non-perturbative effects in the decoupling of the charm
and bottom quarks are negligible in practice. In addition, the effect of neglecting the charm (and thus bottom) quark
contributions to dimensionless ratios of low-energy quantities, which are relevant for the determination of the physical
scale of the lattice simulations (cf.~Section \ref{subsec:HadronicMatching}), has been estimated to be well-below the 
percent level~\cite{Bruno:2014ufa,Knechtli:2017xgy}. It is thus also safe to establish the connection with hadronic
physics using lattice QCD with $\Nf=3$ quark-flavours. Given these observations, we conclude:
\begin{equation}
	\label{eq:AlphaMZ}
	\Lambda^{\Nf=3}_{\overline{\rm MS}}
	\to
	\Lambda^{\Nf=5}_{\overline{\rm MS}}=215(10)(3)\,\MeV
	\quad
	\Rightarrow
	\quad
	\alpha^{(\Nf=5)}_{\overline{\rm MS}}(m_Z) = 0.11852(80)(25).
\end{equation}

The second error in $\Lambda_{\overline{\rm MS}}^{\Nf=5}$, then propagated to $\alpha_s$, comes from 
an estimate (within perturbation theory) of the truncation of the perturbative expansion to 4-loop order 
in computing $\Lambda_{\overline{\rm MS}}^{\Nf=5}/\Lambda_{\overline{\rm MS}}^{\Nf=3}$~\cite{Bruno:2017gxd}. 

\section{Conclusions and Outlook}

In this contribution, we presented a sub-percent precision determination of $\alpha_s$ by means of lattice QCD.
Employing carefully chosen finite-volume couplings and a step-scaling strategy, we overcame the
shortcomings of a naive application of lattice QCD to this problem. This allowed us to keep all  
systematic uncertainties well-under control. We determined the $\Lambda$-parameter of $\Nf=3$ 
QCD by studying the non-perturbative running of the finite-volume couplings from a few hundred MeV
up to O(100 GeV), where the use of perturbation theory was carefully studied. We then computed 
$\Lambda_{\overline{\rm MS}}^{\Nf=5}$ employing a perturbative estimate for $\Lambda_{\overline{\rm MS}}^{\Nf=5}/
\Lambda_{\overline{\rm MS}}^{\Nf=3}$, which has been shown to be accurate within our precision. 
From our result for $\Lambda_{\overline{\rm MS}}^{\Nf=5},$ we finally obtained $\alpha^{(\Nf=5)}_{\overline{\rm MS}}(m_Z)
= 0.11852(84)$, which has an error of $\approx0.7\%$. Our result compares well with the current 
PDG world average: $\alpha^{\text{PDG '18}}_{\overline{\rm MS}}(m_Z)= 0.1181(11)$. 

Studying the error budget of our determination, we can say that the dominant source of uncertainty 
is statistical, and comes from the computation of the non-perturbative running of the couplings. There is 
thus room for improvement in this direction. In addition, one might want to consider including 
explicitly the charm quark in the calculation, in order to rely even less on perturbative information. 
In~conclusion, a determination of $\alpha_s$ with 0.5\% accuracy can be foreseen.
 
\vspace{6pt}

\funding{This research received no external funding.}
\acknowledgments{This work was done as part of the ALPHA Collaboration research programme.  I would like to thank
				 the members of the ALPHA Collaboration and particularly my co-authors of Ref.~\cite{Bruno:2017gxd}
				 for the enjoyable collaboration on this project.}

%

%
\conflictsofinterest{The author declares no conflict of interest.}
%
\reftitle{References}
\externalbibliography{yes}

\end{document}

%% file: macros
\def\rmd{{\rm d}}

\def\proof{\noindent{\sl Proof:}\kern0.6em}

\def\frac#1#2{\hbox{$#1\over#2$}}
\def\dual{\mathstrut^*\kern-0.1em}

\def\lvec#1{\setbox0=\hbox{$#1$}
    \setbox1=\hbox{$\scriptstyle\leftarrow$}
    #1\kern-\wd0\smash{
    \raise\ht0\hbox{$\raise1pt\hbox{$\scriptstyle\leftarrow$}$}}
    \kern-\wd1\kern\wd0}
\def\rvec#1{\setbox0=\hbox{$#1$}
    \setbox1=\hbox{$\scriptstyle\rightarrow$}
    #1\kern-\wd0\smash{
    \raise\ht0\hbox{$\raise1pt\hbox{$\scriptstyle\rightarrow$}$}}
    \kern-\wd1\kern\wd0}
\def\slash#1{\setbox0=\hbox{$#1$}\setbox1=\hbox{$\kern1pt/$}
    #1\kern-\wd0\kern1pt/\kern-\wd1\kern\wd0}

\def\nabstar#1{{\nabla\kern0.5pt\smash{\raise 4.5pt\hbox{$\ast$}}
               \kern-5.5pt_{#1}}}

\def\drvstar#1{{\partial\kern0.5pt\smash{\raise 4.5pt\hbox{$\ast$}}
               \kern-6.0pt_{#1}}}

\def\ldrvstar#1{{\lvec{\,\partial}\kern-0.5pt\smash{\raise 4.5pt\hbox{$\ast$}}
               \kern-5.0pt_{#1}}}

\def\MeV{{\rm MeV}}
\def\GeV{{\rm GeV}}

\def\MSbar{\overline{\rm MS\kern-0.5pt}\kern0.5pt}
\def\Nf{{N_{\rm f}}}

\def\psibar{\overline{\psi}}

\def\zetabar{\bar{\zeta}}
\def\zetaprime{\zeta\kern1pt'}
\def\zetabarprime{\zetabar\kern1pt'}

\def\dirac#1{\gamma_{#1}}
\def\diracstar#1#2{
    \setbox0=\hbox{$\gamma$}\setbox1=\hbox{$\gamma_{#1}$}
    \gamma_{#1}\kern-\wd1\kern\wd0
    \smash{\raise4.5pt\hbox{$\scriptstyle#2$}}}

\def\Obs{{\mathcal O}}

\def\Ds{D_{\rm s}}
\def\DsdagDs{\Ds{\Ds}^{\kern-1pt\dagger}}

\def\avg#1{{\kern1.0pt\overline{\kern-1.0pt#1\kern-1.0pt}\kern1.0pt}}


%% file: universe-406834-eng-done.bbl
\begin{thebibliography}{999}

\bibitem[Salam(2019)]{Salam:2017qdl}
Salam, G.P.
\newblock {The strong coupling: a theoretical perspective}. In {\em From My
  Vast Repertoire ...: Guido Altarelli's Legacy}; Levy, A.; Forte, S.; Ridolfi,
  G., Eds.; World Scientific: Singapore, 2019; pp. 101--121.

\bibitem[d'Enterria(2018)]{dEnterria:2018cye}
d'Enterria, D.
\newblock {$\alpha_s$ status and perspectives (2018)}.
\newblock  In Proceedings of the 26th International Workshop on Deep Inelastic Scattering and
  Related Subjects (DIS 2018) Port Island, Kobe, Japan, 16--20 April 2018.

\bibitem[Tanabashi \em{et~al.}(2018)Tanabashi et~al.]{Tanabashi:2018oca}
Tanabashi, M.; Hagiwara, K.; Hikasa,  K.; Nakamura,  K.;  Sumino,  Y.; Takahashi,  F.; Tanaka,   J.;  Agashe,~K.;  Aielli, G.; Amsler, C.; et al.
\newblock {Review of Particle Physics}.
\newblock {\em Phys. Rev. D} {\bf 2018}, {\em 98},~030001.


\bibitem[Dalla~Brida \em{et~al.}(2016)Dalla~Brida, Fritzsch, Korzec, Ramos,
  Sint, and Sommer]{Brida:2016flw}
Dalla~Brida, M.; Fritzsch, P.; Korzec, T.; Ramos, A.; Sint, S.; Sommer, R.
\newblock {Determination of the QCD $\Lambda$ parameter and the accuracy of
  perturbation theory at high energies}.
\newblock {\em Phys. Rev. Lett.} {\bf 2016}, {\em 117},~182001.

\bibitem[Dalla~Brida \em{et~al.}(2017)Dalla~Brida, Fritzsch, Korzec, Ramos,
  Sint, and Sommer]{DallaBrida:2016kgh}
Dalla~Brida, M.; Fritzsch, P.; Korzec, T.; Ramos, A.; Sint, S.; Sommer, R.
\newblock {Slow running of the Gradient Flow coupling from 200 MeV to 4 GeV in
  $N_{\rm f}=3$ QCD}.
\newblock {\em Phys. Rev. D} {\bf 2017}, {\em 95},~014507.

\bibitem[Bruno \em{et~al.}(2017)Bruno, Dalla~Brida, Fritzsch, Korzec, Ramos,
  Schaefer, Simma, Sint, and Sommer]{Bruno:2017gxd}
Bruno, M.; Dalla~Brida, M.; Fritzsch, P.; Korzec, T.; Ramos, A.; Schaefer, S.;
  Simma, H.; Sint, S.; Sommer, R.
\newblock {QCD Coupling from a Nonperturbative Determination of the
  Three-Flavor $\Lambda$ Parameter}.
\newblock {\em Phys. Rev. Lett.} {\bf 2017}, {\em 119},~102001.

\bibitem[Dalla~Brida \em{et~al.}(2018)Dalla~Brida, Fritzsch, Korzec, Ramos,
  Sint, and Sommer]{DallaBrida:2018rfy}
Dalla~Brida, M.; Fritzsch, P.; Korzec, T.; Ramos, A.; Sint, S.; Sommer, R.
\newblock {A non-perturbative exploration of the high energy regime in
  $N_{\mathrm{f}}=3$ QCD}.
\newblock {\em Eur. Phys. J. C} {\bf 2018}, {\em 78},~372.

\bibitem[Sommer and Wolff(2015)]{Sommer:2015kza}
Sommer, R.; Wolff, U.
\newblock {Non-perturbative computation of the strong coupling constant on the
  lattice}.
\newblock {\em Nucl.~Part. Phys. Proc.} {\bf 2015}, {\em 261--262},~155.

\bibitem[Weinberg(1973)]{Weinberg:1951ss}
Weinberg, S.
\newblock {New approach to the renormalization group}.
\newblock {\em Phys. Rev. D} {\bf 1973}, {\em 8},~3497--3509.

\bibitem[Callan(1970)]{Callan:1970yg}
Callan, Jr., C.G.
\newblock {Broken scale invariance in scalar field theory}.
\newblock {\em Phys. Rev. D} {\bf 1970}, {\em 2},~1541--1547.

\bibitem[Symanzik(1970)]{Symanzik:1970rt}
Symanzik, K.
\newblock {Small distance behavior in field theory and power counting}.
\newblock {\em Commun. Math. Phys.} {\bf 1970}, {\em 18},~227--246.

\bibitem[Symanzik(1971)]{Symanzik:1971vw}
Symanzik, K.
\newblock {Small distance behavior analysis and Wilson expansion}.
\newblock {\em Commun. Math. Phys.} {\bf 1971}, {\em 23},~49.

\bibitem[Montvay and Munster(1997)]{Montvay:1994cy}
Montvay, I.; Munster, G.
\newblock {\em {Quantum Fields on a Lattice}}; Cambridge Monographs on
  Mathematical Physics, Cambridge University Press:  Cambridge, UK, 1997.

\bibitem[Hernandez(2009)]{Hernandez:2009zz}
Hernandez, M.P.
\newblock {Lattice field theory fundamentals}.
\newblock  {Modern perspectives in lattice QCD: Quantum field theory and high
  performance computing. In Proceedings of the International School, 93rd Session, Les~  Houches, France, 3--28 August 2009}; pp. 1--91.

\bibitem[Aoki \em{et~al.}(2017)Aoki et~al.]{Aoki:2016frl}
Aoki, S.; others.
\newblock {Review of lattice results concerning low-energy particle physics}.
\newblock {\em Eur. Phys. J. C} {\bf 2017}, {\em 77},~112,
   doi:10.1140/epjc/s10052-016-4509-7.

\bibitem[L{\"u}scher \em{et~al.}(1991)L{\"u}scher, Weisz, and
  Wolff]{Luscher:1991wu}
L{\"u}scher, M.; Weisz, P.; Wolff, U.
\newblock {A Numerical method to compute the running coupling in asymptotically
  free theories}.
\newblock {\em Nucl. Phys. B} {\bf 1991}, {\em 359},~221.

\bibitem[L{\"u}scher \em{et~al.}(1992)L{\"u}scher, Narayanan, Weisz, and
  Wolff]{Luscher:1992an}
L{\"u}scher, M.; Narayanan, R.; Weisz, P.; Wolff, U.
\newblock {The Schr{\"o}dinger Functional: a renormalizable probe for
  non-abelian gauge theories}.
\newblock {\em Nucl. Phys. B} {\bf 1992}, {\em 384},~168.

\bibitem[Sint(1994)]{Sint:1993un}
Sint, S.
\newblock {On the Schr{\"o}dinger functional in QCD}.
\newblock {\em Nucl. Phys. B} {\bf 1994}, {\em 421},~135.

\bibitem[Sint(1995)]{Sint:1995rb}
Sint, S.
\newblock {One loop renormalization of the QCD Schr{\"o}dinger functional}.
\newblock {\em Nucl. Phys. B} {\bf 1995}, {\em 451},~416--444.

\bibitem[Sint and Sommer(1996)]{Sint:1995ch}
Sint, S.; Sommer, R.
\newblock {The running coupling from the QCD Schr{\"o}dinger functional: a
  one-loop analysis}.
\newblock {\em Nucl.~Phys. B} {\bf 1996}, {\em 465},~71--98.

\bibitem[Narayanan and Wolff(1995)]{Narayanan:1995ex}
Narayanan, R.; Wolff, U.
\newblock {Two loop computation of a running coupling lattice Yang--Mills
  theory}.
\newblock {\em Nucl.~Phys.~B} {\bf 1995}, {\em 444},~425--446.

\bibitem[Bode \em{et~al.}(1999)Bode, Wolff, and Weisz]{Bode:1998hd}
Bode, A.; Wolff, U.; Weisz, P.
\newblock {Two-loop computation of the Schr{\"o}dinger functional in pure SU(3)
  lattice gauge theory}.
\newblock {\em Nucl.~Phys.~B} {\bf 1999}, {\em 540},~491--499.

\bibitem[Bode \em{et~al.}(2000)Bode, Weisz, and Wolff]{Bode:1999sm}
Bode, A.; Weisz, P.; Wolff, U.
\newblock {Two-loop computation of the Schr{\"o}dinger functional in lattice
  QCD }.
\newblock {\em Nucl.~Phys.~B} {\bf 2000}, {\em 576},~517--539.

\bibitem[Dalla~Brida and Hesse(2014)]{Brida:2013mva}
Dalla~Brida, M.; Hesse, D.
\newblock {Numerical Stochastic Perturbation Theory and the Gradient Flow}.
\newblock {\em PoS} {\bf 2014}, {\em Lattice2013},~326.

\bibitem[Dalla~Brida and Lüscher(2016)]{DallaBrida:2016dai}
Dalla~Brida, M.; Lüscher, M.
\newblock {The gradient flow coupling from numerical stochastic perturbation
  theory}.
\newblock {\em PoS} {\bf 2016}, {\em LATTICE2016},~332.

\bibitem[Dalla~Brida and Lüscher(2017)]{DallaBrida:2017tru}
Dalla~Brida, M.; Lüscher, M.
\newblock {SMD-based numerical stochastic perturbation theory}.
\newblock {\em Eur. Phys. J. C} {\bf 2017}, {\em 77},~308.

\bibitem[L{\"u}scher \em{et~al.}(1994)L{\"u}scher, Sommer, Weisz, and
  Wolff]{Luscher:1993gh}
L{\"u}scher, M.; Sommer, R.; Weisz, P.; Wolff, U.
\newblock {A precise determination of the running coupling in the SU(3)
  Yang--Mills theory}.
\newblock {\em Nucl. Phys. B} {\bf 1994}, {\em 413},~481--502.

\bibitem[de~Divitiis \em{et~al.}(1995)de~Divitiis, Frezzotti, Guagnelli,
  L{\"u}scher, Petronzio, Sommer, Weisz, and Wolff]{deDivitiis:1994yz}
de~Divitiis, G.; Frezzotti, R.; Guagnelli, M.; L{\"u}scher, M.; Petronzio, R.;
  Sommer, R.; Weisz, P.; Wolff, U.
\newblock {Universality and the approach to the continuum limit in lattice
  gauge theory}.
\newblock {\em Nucl. Phys.} {\bf 1995}, {\em B437},~447--470.

\bibitem[Narayanan and Neuberger(2006)]{Narayanan:2006rf}
Narayanan, R.; Neuberger, H.
\newblock {Infinite $N$ phase transitions in continuum Wilson loop operators}.
\newblock {\em JHEP} {\bf 2006}, {\em 03},~064.

\bibitem[L{\"u}scher(2010)]{Luscher:2010iy}
L{\"u}scher, M.
\newblock {Properties and uses of the Wilson flow in lattice QCD}.
\newblock {\em J. High Energy Phys.} {\bf 2010}, {\em 2010},~071.

\bibitem[L{\"u}scher and Weisz(2011)]{Luscher:2011bx}
L{\"u}scher, M.; Weisz, P.
\newblock {Perturbative analysis of the gradient flow in non-abelian gauge
  theories}.
\newblock {\em J. High Energy Phys.} {\bf 2011}, {\em 2011},~51.

\bibitem[Fodor \em{et~al.}(2012)Fodor, Holland, Kuti, Nogradi, and
  Wong]{Fodor:2012td}
Fodor, Z.; Holland, K.; Kuti, J.; Nogradi, D.; Wong, C.H.
\newblock {The Yang--Mills gradient flow in finite volume}.
\newblock {\em J. High Energy Phys.} {\bf 2012}, {\em 2012},~7.

\bibitem[Fritzsch and Ramos(2013)]{Fritzsch:2013je}
Fritzsch, P.; Ramos, A.
\newblock {The gradient flow coupling in the Schr\"odinger Functional}.
\newblock {\em J. High Energy Phys.} {\bf 2013}, {\em 2013},~8.

\bibitem[Ramos and Sint(2016)]{Ramos:2015baa}
Ramos, A.; Sint, S.
\newblock {Symanzik improvement of the gradient flow in lattice gauge
  theories}.
\newblock {\em Eur. Phys. J.} {\bf 2016}, {\em C76},~15.

\bibitem[Fritzsch \em{et~al.}(2014)Fritzsch, Dalla~Brida, Korzec, Ramos, Sint,
  and Sommer]{Brida:2014joa}
Fritzsch, P.; Dalla~Brida, M.; Korzec, T.; Ramos, A.; Sint, S.; Sommer, R.
\newblock {Towards a new determination of the QCD Lambda parameter from running
  couplings in the three-flavour theory}.
\newblock {\em PoS} {\bf 2014}, {\em LATTICE2014},~291.

\bibitem[Sommer(2014)]{Sommer:2014mea}
Sommer, R.
\newblock {Scale setting in lattice QCD}.
\newblock {\em PoS} {\bf 2014}, {\em LATTICE2013},~015.

\bibitem[Bruno \em{et~al.}(2017)Bruno, Korzec, and Schaefer]{Bruno:2016plf}
Bruno, M.; Korzec, T.; Schaefer, S.
\newblock {Setting the scale for the CLS 2+1 flavor ensembles}.
\newblock {\em Phys. Rev.} {\bf 2017}, {\em D95},~74504.

\bibitem[Bruno \em{et~al.}(2015)Bruno et~al.]{Bruno:2014jqa}
Bruno, M.; others.
\newblock {Simulation of QCD with $N_{\rm f} = 2 + 1$ flavors of
  non-perturbatively improved Wilson fermions}.
\newblock {\em J. High Energy Phys.} {\bf 2015}, {\em 2015},~43.

\bibitem[Dalla~Brida \em{et~al.}(2016)Dalla~Brida, Sint, and
  Vilaseca]{Brida:2016rmy}
Dalla~Brida, M.; Sint, S.; Vilaseca, P.
\newblock {The chirally rotated Schrödinger functional: theoretical
  expectations and perturbative tests}.
\newblock {\em J. High Energy Phys.} {\bf 2016}, {\em 2016},~102.

\bibitem[Dalla~Brida \em{et~al.}(2018)Dalla~Brida, Korzec, Sint, and
  Vilaseca]{DallaBrida:2018tpn}
Dalla~Brida, M.; Korzec, T.; Sint, S.; Vilaseca, P.
\newblock {High precision renormalization of the flavour non-singlet Noether
  currents in lattice QCD with Wilson quarks}.  \emph{arXiv} {\bf 2018},
 arXiv:1809.03383.

\bibitem[Athenodorou \em{et~al.}(2018)Athenodorou, Finkenrath, Knechtli,
  Korzec, Leder, Marinković, and Sommer]{Athenodorou:2018wpk}
Athenodorou, A.; Finkenrath, J.; Knechtli, F.; Korzec, T.; Leder, B.;
  Marinković, M.K.; Sommer, R.
\newblock {How perturbative are heavy sea quarks?} \emph{arXiv} {\bf 2018},
\newblock  arXiv:1809.03383.

\bibitem[Bruno \em{et~al.}(2015)Bruno, Finkenrath, Knechtli, Leder, and
  Sommer]{Bruno:2014ufa}
Bruno, M.; Finkenrath, J.; Knechtli, F.; Leder, B.; Sommer, R.
\newblock {Effects of heavy sea quarks at low energies}.
\newblock {\em Phys. Rev. Lett.} {\bf 2015}, {\em 114},~102001.

\bibitem[Knechtli \em{et~al.}(2017)Knechtli, Korzec, Leder, and
  Moir]{Knechtli:2017xgy}
Knechtli, F.; Korzec, T.; Leder, B.; Moir, G.
\newblock {Power corrections from decoupling of the charm quark}.
\newblock {\em Phys. Lett. B} {\bf 2017}, {\em 774},~649--655.

\end{thebibliography}
